\documentclass[PRB,twocolumn]{revtex4}
\usepackage{}
\usepackage{epsfig}
\usepackage{graphicx}
\usepackage{amsfonts,amssymb}
\usepackage{amsmath}
\usepackage{color}
\usepackage {times}
\usepackage{lipsum}
\usepackage{bm}
\usepackage{blindtext}
\usepackage{sidecap}
\usepackage{floatrow}
\usepackage{mwe}
\usepackage{subfigure}
\usepackage{nccmath}
\usepackage[title]{appendix} 
\usepackage[utf8]{inputenc}
\usepackage[T1]{fontenc}
\DeclareUnicodeCharacter{0308}{}
\usepackage{mathptmx} 
\definecolor{refcolor}{rgb}{1.0,0.0,0.0}

\newcommand{\be}{\begin{equation}}
\newcommand{\ee}{\end{equation}}   
\newcommand{\bea}{\begin{eqnarray}}
\newcommand{\eea}{\end{eqnarray}}
\newcommand{\ba}{\begin{array}}
\newcommand{\ea}{\end{array}}

\begin{document}

\author{Garima Goyal and Dheeraj Kumar Singh}
\email{dheeraj.kumar@thapar.edu }
\affiliation{ School of Physics and Materials Science, Thapar Institute of Engineering and Technology, Patiala-147004, Punjab, India}     

\date{\today}

\title{Semimetallic spin-density wave state in iron pnictides}

\begin{abstract}
   
We examine the existence of semimetallic spin-density wave states in iron pnictides. In the experimentally observed metallic spin-density wave state, the symmetry-protected Dirac cones are located away from the Fermi surface giving rise to tiny pockets and there are also additional Fermi pockets such as one around $\Gamma$. We find that the location of a pair of Dirac points with respect to the Fermi surface exhibits significant sensitivity to the orbital splitting between the $d_{xz}$ and $d_{yz}$ orbitals. Besides, in the presence of orbital splitting, the Fermi pockets not associated with the Dirac cones, can be suppressed so that a semimetallic spin-density wave state can be realized. We explain these finding in terms of difference in the slopes and orbital contents of the bands which form the Dirac cone, and obtain the necessary conditions dependent on these two and other parameters for the coexisting Dirac semimetallic and spin-density wave states. Additionally, the topologically protected edge states are studied in the ribbon geometry when the same are oriented either along $x$ or $y$ axes. 
\end{abstract}

\maketitle
\section{Introduction} 
Iron-based multiband superconductors have attracted considerable attention in recent times because of their complex band structure~\cite{anderson, burkov} and variety of phases they can exhibit including unconventional superconductivity, nematic order and other novel phases~\cite{kordyuk, stewart}. However, a renewed interest has been generated largely because of the signatures of topological states obtained in this class of superconducting materials~\cite{zhang, zhang_p}. 

Zero-energy Majorana bound states were found to exist on the surface of superconducting iron chalcogenides as indicated by the scanning-tunneling microscopy (STM)~\cite{wang, liu*}. These zero-energy bound states were attributed to the band inversion between the bands dominated by the pnictogen $p_z$ and iron $d_{xz} /d_{yz}$ orbitals~\cite{zhang,wang1, wu}. For similar reasons, it was suggested that the iron-based superconductors (IBS) could exhibit a band topology similar to that of a topologically insulator (TI) or Dirac semimetal (DS), which was substantiated by the laser-based spin- and angle-resolved photoemission spectroscopies (ARPES)~\cite{zhang}. 

\begin{figure}
    \centering
    \begin{minipage}{0.5\textwidth}
        \centering
        \includegraphics[width=0.92\textwidth]{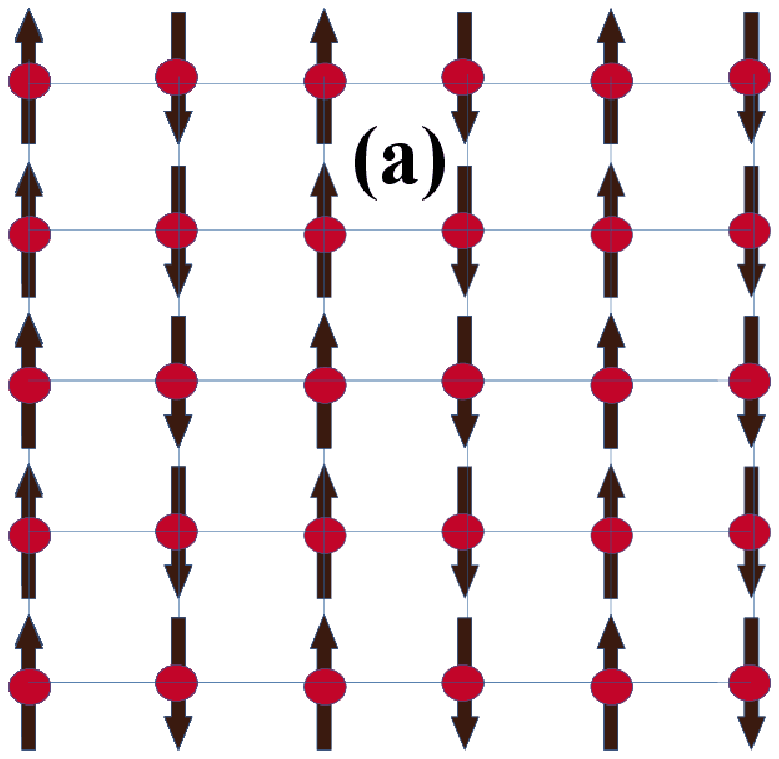} 
    \end{minipage}\hfill
    \begin{minipage}{0.5\textwidth}
        \centering
        \includegraphics[width=0.92\textwidth, angle=-90]{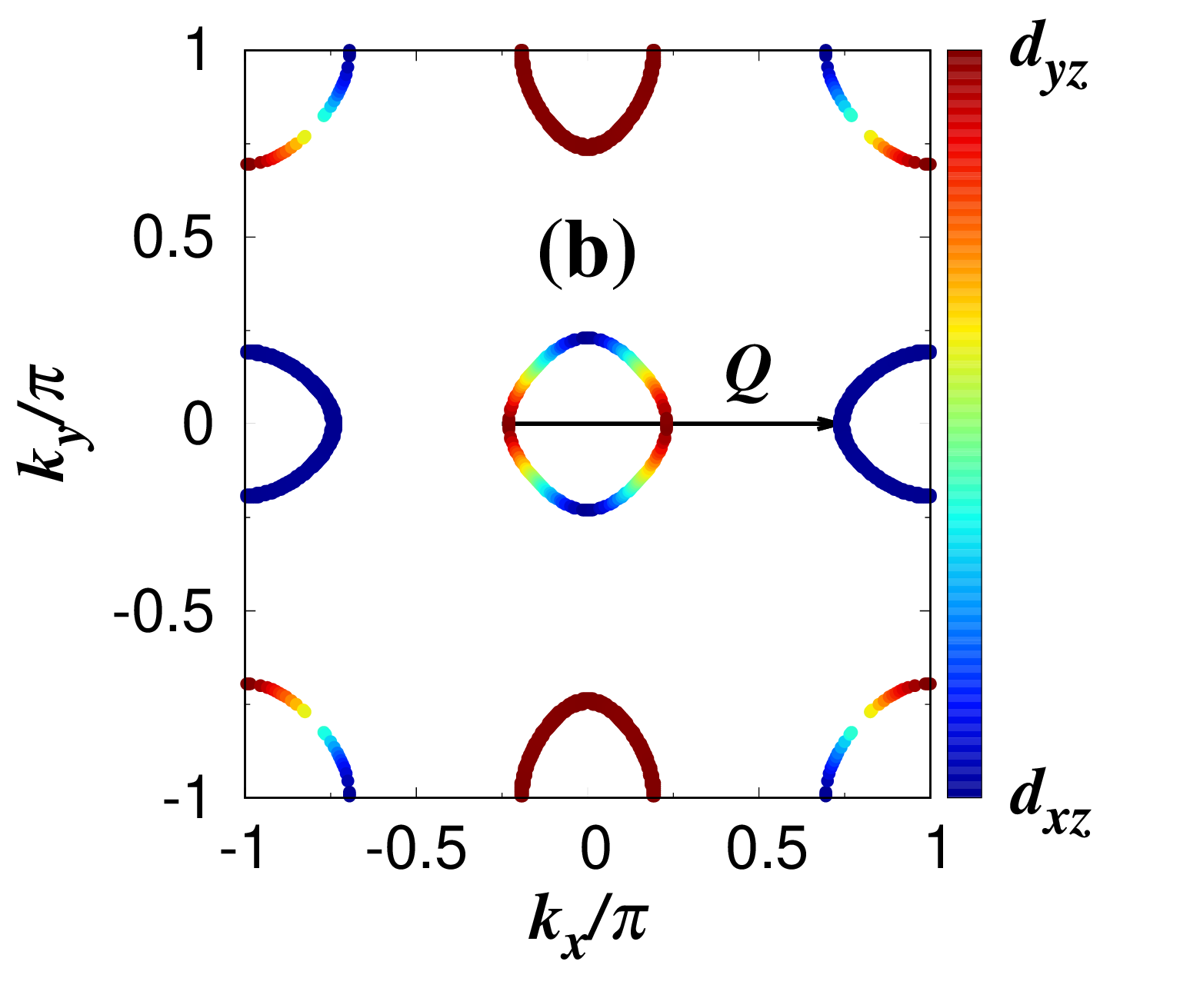} 
    \end{minipage}
\caption{(a) Schematic representation of spin arrangements in the $(\pi,0)$ SDW state. (b) Fermi surfaces in the unordered state within the two-orbital model. The nesting vector ${\bf Q} = (\pi,0)$ connects the hole pocket at $\Gamma = (0,0)$ to the electron pocket at $X = (\pi,0)$. The color palette shows the variation of orbital-charge densities for the $d_{xz}$ and $d_{yz}$ orbitals.}
 \label{1}
\end{figure}

Evidence of Dirac cone in the four-fold rotational symmetry broken metallic spin-density wave (SDW) state have been obtained through ARPES~\cite{richard} and quantum oscillation measurements~\cite{sutherland}, which was predicated to be gapless~\cite{ran}. The SDW state, with a collinear or striped magnetic order, consists of chains of magnetic moments pointing along the same direction while the moments of the neighboring chains are aligned along the opposite directions~\cite{lumsden, johnston} [Fig.~\ref{1}(a)]. This state, with an ordering wavevector ($\pi, 0$), is considered widely to be a consequence of the inherent Fermi-surface instability as there exists a very good nesting in between the hole- and electron-pockets around $\Gamma $ at $(0, 0)$ and X at ($\pi, 0$) points, respectively~\cite{korshunov, raghu, graser, brydon}. 

The gaplessness of the SDW state arises from  the symmetry and band topology~\cite{ran}. In the unordered state, the band touching around (0, 0) (and around an additional point ($\pi, \pi$) in the two-orbital model) leads to vorticities equal to $\pm$2 for the hole pockets. The vorticity corresponds to a complete rotation of the spinor defined in terms of two orbitals $d_{xz}$ and $d_{yz}$, which is also reflected in the contribution of different orbitals to the hole pocket [Fig. \ref{1}(b)]. It vanishes, on the other hand, for the electron pockets around ($\pi, 0$) or ($0, \pi$) as the pockets are dominated by a single orbital. As a result, there is a vorticity mismatch for the hole and electron pockets, which subsequently leads to a gapless SDW state accompanied with an even number of Dirac cones, despite the presence of a good nesting.  

The robustness of the Dirac cones and nodes originates from three symmetries associated with the metallic SDW state: collinearity of the magnetic order, inversion symmetry about an iron atom and another symmetry which combines together the time reversal and inversion of magnetic moments~\cite{ran}. These cones in the SDW state are not far away from the Fermi level~\cite{richard}. Thus, there is a strong possibility of obtaining a coexisting DS and SDW states by tuning parameters accessible through experiments. A DS state is characterized by a linear band crossing of conduction and valence bands at the Fermi level~\cite{castro, hasan1}. The band-crossing point, also known as Dirac point (DP), is four-fold degenerate. The massless fermion in the vicinity of the DP, with several novel electronic behavior~\cite{li, wang3, lundgren}, is described by the Dirac equation~\cite{castro,hasan1}. 

The search for materials or phases hosting DS besides graphene, which turns into a topological insulator (TI) because of a small spin-orbit couplings, has gathered momentum lately~\cite{chowdhury}. Potential existence of DS has been indicated in 3D compounds such as $\beta$-cristobalite BiO$_2$~\cite{young}, BiZnSiO$_4$~\cite{steinberg}, Cd$_3$As$_2$~\cite{liu1, wang5, borisenko} etc. and efforts are being made to explore this state in new 2D systems as well~\cite{ni,wang4}. 

In this paper, we investigate coexisting DS and SDW states of iron pnictides. Among several parameters whose variation do not affect the symmetries required for the stable Dirac cones, we show that the location of DPs with respect to the Fermi surface can be tuned by the orbital splitting between the $d_{xz}$ and $d_{yz}$ orbitals in addition to the band filling. Experimentally, while the latter can be controlled by charge carrier doping, the former can be achieved by applying in-plane stress on the sample. Another consequence of orbital splitting is that the additional Fermi pockets, apart from the ones associated with the Dirac cones, may disappear so that the resulting state has only Dirac cones crossing the Fermi level leading to tiny Fermi pockets or DPs. In the latter case, the state is semimetallic SDW state, for which we calculate necessary conditions dependent on orbital splitting and other parameters. In addition, linearlized dispersions are obtained near the Dirac points and the nature of associated edge states is also studied.


\section{Model and method}
Iron pnictides have a quasi-two dimensional crystal structure where the square lattices formed by Fe and As atoms are interlaced together so that the As atoms are positioned a little above and below the center of each square plane of Fe atoms. Therefore, a crystallographic unit cell consisting of two Fe and two As atoms is formed. The Fe 3$d$ orbitals hybridize with the As 4$p$ orbitals, which increases the complexity of band structure. Bandstructure calculations indicate that the major contributions to the density of state (DOS) at the Fermi level comes from $d_{xz}$, $d_{yz}$ and $d_{xy}$ orbitals~\cite{graser}. In the following, we consider the tight-binding Hamiltonian based on two orbitals $d_{xz}$ and $d_{yz}$, which is given by \cite{raghu}
\begin{equation}
 H_k  =  \sum_{ \bf{ij}} \sum_{\mu ,\nu} \sum_{\sigma} t_{\bf{ij}}^{\mu \nu} d_{\bf{i} \mu \sigma}^{\dagger}  d^{}_{\bf{j} \nu \sigma}  -  \delta \sum_{\bf{i}, \sigma} (d^{\dagger}_{{\bf i} xz \sigma}  d^{}_{{\bf i} xz \sigma}  - d_{{\bf i} yz \sigma}^{\dagger} d^{}_{{\bf i} yz \sigma}),
 \end{equation}
 where $t_{\bf{ij}}^{\mu \nu}$ are the intra- and inter-orbital hopping parameters for the first and second nearest neighbor. $\textit{d}_{\bf{i} \mu \sigma}^{\dagger} (\textit{d}_{\bf{i} \mu \sigma})$ denotes the creation (annihilation) operators for the orbital $\mu$ at site ${\bf i}$ and spin $\sigma$. The second term with parameter $\delta$ takes into account the orbital splitting (OS) between the two orbitals, which has been observed to exist up to a very high temperature even beyond the SDW transition temperature~\cite{Yi,kasahara1}.  

The standard on site Coulomb interaction terms are
 \begin{eqnarray}
 H_{i}  &=&  U \sum_{{\bf i}  \mu} n_{\bf{i} \mu \uparrow} n_{\bf{i}  \mu \downarrow}  +  \left(U^{\prime} -  \frac{J}{2}\right) \sum_{\bf{i}} n_{\bf{i} \mu} n_{\bf{i} \nu} \nonumber \\
 &-&  2J \sum_{\bf{i}}{\bf S}_{i \mu} \cdot {\bf S}_{i \nu}  +  J\sum_{\bf{i},  \sigma} d^{\dagger}_{{\bf i}\mu \sigma} d^{\dagger}_{{\bf i}\mu \bar{\sigma}}d_{{\bf i}\nu \bar{\sigma}}d_{{\bf i}\nu \sigma} 
 \end{eqnarray}
 The first and second terms describe the intra- and inter-orbital Coulomb interaction, respectively, where $n_{{\bf i}\mu \sigma} = d^{\dagger}_{{\bf i} \mu \sigma} d^{}_{{\bf i} \mu      \sigma}$ and $n_{{\bf i}\mu} = \sum_{\sigma {\sigma}^{\prime}}d^{\dagger}_{{\bf i} \mu \sigma} d^{}_{{\bf i} \mu \sigma}$. The third and fourth terms take into  account the Hund’s coupling and the pair hopping, where ${S}^{j}_{i \mu}= \sum_{\sigma}d^{\dagger}_{{\bf i} \mu \sigma} \sigma^{j}_{\sigma {\sigma}^{\prime}} d^{}_{{\bf i} \mu {\sigma}^{\prime}}$ and $\bar{\sigma}$ denotes the spin anti-parallel to $\sigma$. The relation $\textit{U}   =  U^{\prime}  -  2J$ is ensured to keep the rotational symmetry intact.
   
Various interaction terms in the Hamiltonian can be meanfield decoupled in the SDW state, which yield terms bilinear in the electron creation (or annihilation) operator for the SDW state. These terms, when the magnetic moments are oriented along $z$ direction, are  
\begin{eqnarray}
{H}_{\rm mf}^{i} &=& \frac{U}{2} \sum_{{\bf k} \mu \sigma} \bigg(- s \sigma  m_{\mu}  + n_{\mu}\bigg) d_{{\bf k}\mu\sigma}^\dagger d_{{\bf k}\mu\sigma} \nonumber\\ &+& (U' - \frac{J}{2}) \sum_{{\bf k}, \mu \neq \nu, \sigma}  n_{\nu}  d_{{\bf k}\mu\sigma}^\dagger d_{{\bf k}\mu\sigma} \nonumber\\ &-& \frac{J}{2} \sum_{{\bf k}, \mu \neq \nu, \sigma} \sigma s m_{\nu} d_{{\bf k}\mu\sigma}^\dagger d_{{\bf k}\mu\sigma}.
\end{eqnarray}
$n_{\mu}$ and $m_{\mu}$ are the orbital-resolved  charge density and sublattice magnetization for the orbital $\mu$, respectively. \textit{s} and $\sigma$ are equal to 1 (-1) for A (B) sublattice and $\uparrow$ spin ($\downarrow$ spin), respectively. 

After combining the kinetic, OS, and meanfield decoupled interaction parts, the Hamiltonian for the ($\pi$, 0) SDW state in two sublattice basis is \\
 \begin{equation}
     \mathcal{H}_{mf} = \sum_{{\bf k} \sigma} \Psi_{{\bf k} \sigma}^{\dagger} (\hat{T}_{{\bf k} \sigma} + \hat{M}_{{\bf k} \sigma}) \Psi_{{\bf k} \sigma},
 \end{equation}
 where the matrix elements $\textit{T}_{{\bf k} \sigma}^{\mu \nu}$ and $\textit{M}_{{\bf k} \sigma}^{\mu \nu} = - s \Delta_{\mu \nu} + \frac{5 \textit{J} - \textit{U}}{2} \textit{n}_{\mu} \delta^{\mu \nu}$ are the kinetic and interaction part contributions. $s$ is $+/-$ on $A/B$ sublattice. The electron-field operator is $\Psi_{k \sigma}^{\dagger } = ( \textit{d}_{A {\bf k}1 \sigma}^{\dagger}, \textit{d}_{A {\bf k}2 \sigma}^{\dagger}, \cdots \textit{d}_{B {\bf k}1 \sigma}^{\dagger}, \textit{d}_{B {\bf k}2 \sigma}^{\dagger},  \cdots )$ and the exchange fields is $2 \Delta_{\mu \nu} = \textit{U m}_{\mu} \delta^{\mu \nu}  + \textit{J} \delta^{\mu \nu} \sum_{\mu \neq \nu} \textit{m}_{\nu}$. The charge density $\textit{n}_{\mu}$ and magnetization $\textit{m}_{\mu}$ are calculated self consistently by diagonalizing the Hamiltonian matrix present in Eq. 4.  
  
\section{Two-orbital model}  

In the two-orbital model, the meanfield SDW Hamiltonian in the two sublattice basis $A$ with spin up and $B$ with spin down is given by
\begin{equation}
H_{\rm MF} = \sum_{\bf{k},\sigma} \Psi^{\dagger}_{\bf{k} \sigma}
\begin{pmatrix}
 H^{\sigma}_{\alpha \alpha}({\bf k})  &  H^{\sigma}_{\alpha \beta}({\bf k}) \\
 H^{\sigma \dagger}_{\alpha \beta}({\bf k})  &  H^{\sigma}_{\beta \beta}({\bf k}) \\ 
\end{pmatrix} \Psi^{}_{\bf{k}\sigma}
\end{equation}
with $\Psi^{\dagger}_{\bf{k} \sigma} = (d^{\dagger}_ {A\alpha \sigma}$,$d^{\dagger}_{B\alpha \sigma}$,$d^{\dagger}_{A\beta \sigma}$,$d^{\dagger}_{B\beta \sigma})$, where the sub- or super-script $\alpha$ and $\beta$ are used to denote the orbitals $d_{xz}$ and $d_{yz}$, respectively, throughout. The matrices in Eq. 5 are 
\begin{equation}
 H^{\sigma}_{\alpha \alpha}({\bf k}) =
    \begin{pmatrix}
    \epsilon^{\alpha \alpha}_y - \sigma \Delta_{\alpha } - \delta + N_{\alpha}   &  \epsilon^{\alpha \alpha}_x + \epsilon^{\alpha \alpha}_{xy} \\
    \epsilon^{\alpha \alpha}_x + \epsilon^{\alpha \alpha}_{xy}  &  \epsilon^{\alpha \alpha}_y + \sigma \Delta_{\alpha} - \delta + N_{\alpha}  \\
    \end{pmatrix}
\end{equation}
and
\begin{equation}
 H_{\alpha \beta}({\bf k}) =
    \begin{pmatrix}
    0  &  \epsilon^{\alpha \beta}_{xy} \\
    \epsilon^{\alpha \beta}_{xy}  &  0 \\
    \end{pmatrix}.
\end{equation}
The intra- and inter-orbital hopping parameters along $x$ and $y$ directions for different orbitals are given by 
\begin{eqnarray}
\epsilon^{\alpha \alpha}_x &=& -2t_1 \cos k_x, \,\,\, \epsilon^{\beta \beta}_x = -2t_2 \cos k_x \nonumber\\
\epsilon^{\beta \beta}_y &=& -2t_1 \cos k_y, \,\, \epsilon^{\alpha \alpha}_y = -2t_2 \cos k_y   \nonumber\\
\epsilon^{\alpha \alpha}_{xy} &=& \epsilon^{\beta \beta}_{xy} = -4t_3 \cos k_x \cos k_y\nonumber\\
\epsilon^{\alpha \beta}_{xy} &=& -4t_4 \sin k_x \sin k_y 
\end{eqnarray}
The hopping parameters $t_1$ and $t_2$ link similar orbitals corresponding to the nearest neighbor $\sigma$ and $\pi$ bonds respectively, while the the next-nearest neighbour hopping parameters $t_3$ and $t_4$ denote the overlap amplitude between two similar and dissimilar orbitals, respectively. Various hopping parameters considered for our calculations are $t_1$ = -1.0, $t_2$ = 1.3, $t_3$ = $t_4$ = -0.85~\cite{raghu}. Herefrom, we set $|t_1|$ as the unit of energy. The exchange field $\Delta_{\alpha / \beta}$ and charge-density dependent $N_{\alpha / \beta}$ are given by 
\begin{equation}
\Delta_{\alpha / \beta} = (Um_{\alpha / \beta} + J m_{\beta / \alpha})/2, \,\,\,
N_{\alpha / \beta} = (5J-U) n_{\alpha / \beta}/2
\end{equation}
where $m_{\alpha / \beta}$ and $n_{\alpha / \beta}$ are magnetization and charge density for $d_{xz/yz}$ orbital, respectively.

We will first examine the parameter space in the theory of SDW state to search for the DS state coexisting with SDW state. All the on-site Coulomb interaction parameters are fixed unless stated otherwise. We have chosen $U/|t_1|\sim $ 4 or $U\sim W/3$ with $W$ being the bandwidth~\cite{yang} and $J\sim 0.18U$ nearly in the middle of the range $0.15U \lesssim J \lesssim 0.25U$ in accordance with various  estimates~\cite{miyake,ishida}. The chemical potential is fixed throughout so that the band filling is $n = 2$. 

\subsection{Bulk dispersion}
\begin{figure}
       \centering
       \includegraphics[scale=1.0, width=8.7cm]{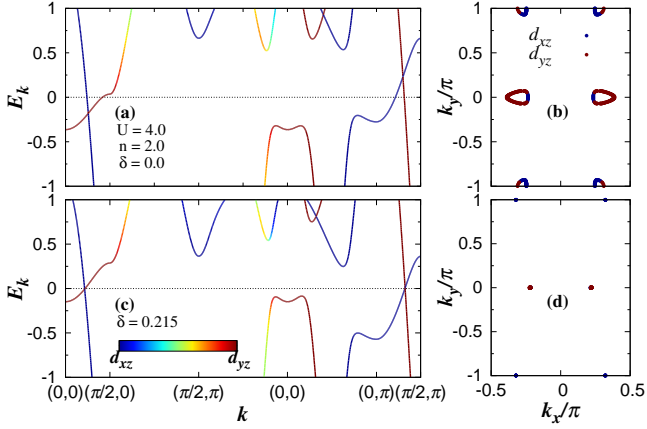}
       \vspace{0mm}
    \caption{Energy bands in the SDW state with $U = 4$ and $J = 0.18U$ when the OS parameters are (a) $\delta = 0.0$ and (c) $\delta = 0.215$. The varying color scheme used for the dispersion curve represents orbital-charge density. Corresponding Fermi surfaces are shown in figures (b) and (d), respectively. The same color scheme also shows the dominating orbital along the Fermi surfaces.}
    \label{2}
\end{figure}

Fig.~\ref{2} shows the quasiparticle dispersion and Fermi surface in the SDW state for different values of OS parameter $\delta$. For $\delta = 0$ [Fig.~\ref{2}(a)], there exist two pairs of Dirac cones $D_1$ and $D_2$, $D_1$ along the $\Gamma$-X and $D_2$ along  ${\rm X^{\prime}}$-M directions. Former is below the Fermi level while the latter one is above it. The orbital distribution of the Fermi pockets are largely similar for both the pair of Dirac cones. As noticed, the face of the pockets towards $k_y = 0$ is dominated by $d_{xz}$ orbital whereas by $d_{yz}$ orbital towards $k_y = \pm \pi/2$ [Fig.~\ref{2}(b)]. 

One of the parameters which can control the location of Dirac cones with respect to the Fermi level is the band filling. However, it will shift both the Dirac cones either up or down together so that the Fermi pocket associated with one of the cones with increase in size while the other will decrease. Thus, the DS-SDW state cannot be obtained by merely doping charge carriers. On the other hand, we find that by increasing OS parameter $\delta$, $D_1$ and $D_2$ simultaneously can be pushed up and down, respectively, so that both the associated DPs may approach the Fermi level together, as required to achieve the coexisting DS and SDW states [Fig.~\ref{2}(c)]. Fig. 2(d) shows the Fermi pockets obtained for $\delta \sim 0.2$, where both DPs associated with $D_1$ and $D_2$ can be seen at the Fermi level as indicated by the Fermi pockets turning into the Fermi points in the reduced-Brillouin zone. 

The shifting up or down of the DPs can be understood with the help of slope of the bands dominated either by $d_{xz}$ orbital or by $d_{yz}$ orbital, which cross each other to generate the DPs. For $D_1$, the ratio $r$ of the absolute value of the slope of the bands dominated by $d_{xz}$ orbital and $d_{yz}$ orbital is $r>1$. Thus, a positive $\delta$, which lowers the energy of $d_{xz}$ orbital and elevate the energy of $d_{yz}$ orbital, will shift $D_1$ upward. Similarly, it may be noted that $r<1$ for $D_2$. Therefore, when the energy of $d_{xz}$ orbital is lowered, the band dominated by $d_{xz}$ orbital will shift down, which will, in turn, bring down $D_2$. Both $D_1$ and $D_2$ approach Fermi level when the splitting $\delta$ increases. In this way, the location of Dirac points can be controlled with the help of OS parameter $\delta$.

\begin{figure}
    \centering
    \includegraphics[scale=1.0, width=8.7cm]{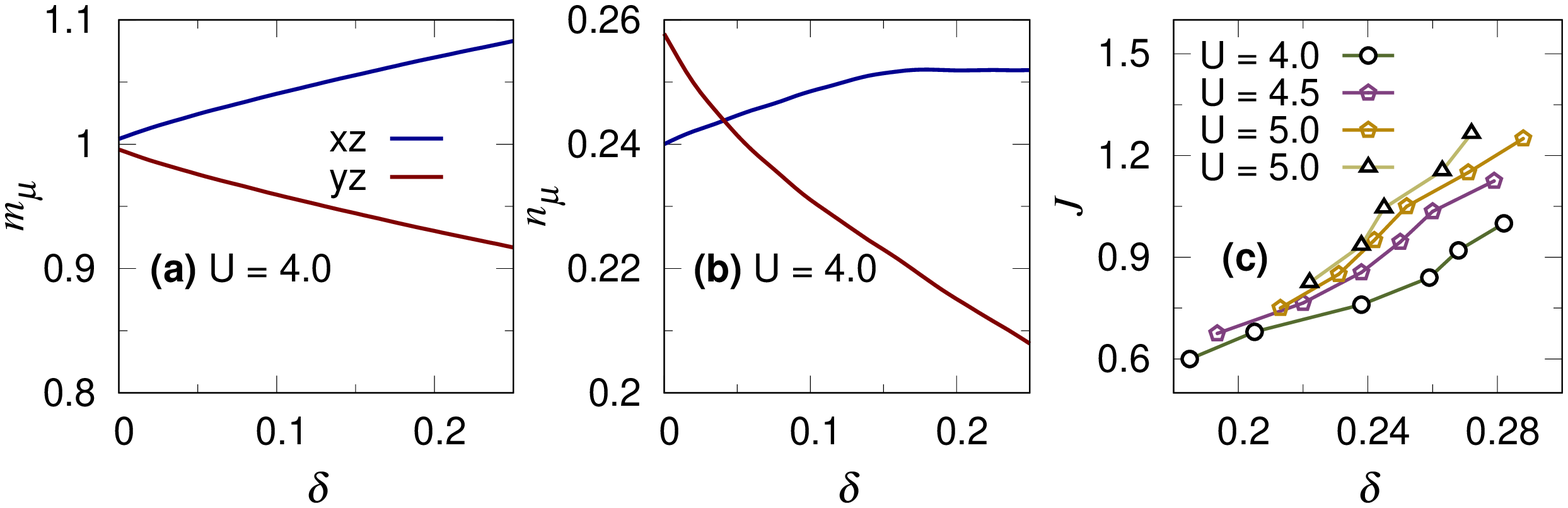}
    \vspace{-4mm}
    \caption{(a) Orbital magnetization and (b) charge density as functions of OS parameter $\delta$ between $d_{xz}$ and $d_{yz}$ orbitals in the ordered state for $U = 4$ and $J = 0.18U$. (c) Self-consistently obtained DS-SDW states for various $U$s in $J$-$\delta$ space for band filling $n = 2$.}
    \label{3}
\end{figure}

Fig.~\ref{3} shows self-consistently obtained orbital-resolved magnetization and charge density as functions of OS parameter $\delta$. As expected, the $d_{xz}$ and $d_{yz}$ orbital charge density increases and decreases with $\delta$, respectively. However, the behavior of magnetization is in contrast with what is expected conventionally. When the orbital filling continues to increase beyond unity, the magnetization is expected to decrease. Similarly, when the orbital filling continues to drop below unity then the magnetization is expected to rise. On the contrary, one notices that the $d_{xz}$ orbital magnetization $m_{xz}$ first increases slightly and then becomes nearly constant whereas $m_{yz}$ shows a relatively sharper decline. This may arise because of a subtle interplay between bandstructure and correlation effect, as the largest interaction $U \sim 4$ is in the intermediate coupling regime. 

The self-consistently obtained curves denoting the existence of DS-SDW state for different $U$s are shown in the $J$-$\delta$ space [Fig. \ref{3}(c)]. It may be noted that for all $U$s considered, $J$ increases with $\delta$ within the range $0.15U \lesssim  J \lesssim 0.25U$. The range of $J$ is dependent on $U$, for this reason, $J$ is in the unit of $|t_1|$ in the plot. We note that the range of $\delta$ shrinks as $U$ increases, which implies more sensitivity to any change in $\delta$ for higher $U$. However, for $U = 4$ a value close to various estimates, we find a relatively broad range of $\delta$ with value extending from $\delta=$ 0.18 to 0.27.

\subsection{DS-SDW conditions} In order to obtain the DS-SDW state, the self-consistency should be achieved subjected to conditions, which will be discussed below. First of all, we focus along the direction $k_y=0 \,\, {\rm or} \,\, \pi$ in the Brillouin zone, where $H_{\alpha \beta}$ becomes a null matrix. Therefore, $H_{\rm MF}$ takes a block diagonal form with two block Hamiltonians $H_{\alpha \alpha} ({\bf k})$ and $H_{\beta \beta} ({\bf k})$ corresponding to each orbital with size 2 $\times$ 2 matrices, which can be readily diagonalized. The energy eigenvalues are given by
    \begin{equation}
    E^x_{s {\pm}} = (\epsilon^{ss}_ y \pm \delta + N_{s} - \mu) \pm \sqrt{(\epsilon^{ss}_x + \epsilon^{ss}_{xy})^{2} + \Delta_{s}^{2}} 
    \end{equation}
 with setting $k_y = 0$ or $\pi$. $s$ refers to $\alpha$ or $\beta$ orbitals. Superscript $x$ denotes the fact that the energy is essentially a function of $k_x$ only for $k_y = 0$ or $\pi$, while subscript $s$ merely indicates that the dispersion is dominated by either of the orbitals. Two dispersions $E^x_{\alpha +}$ and $E^x_{\beta -}$ cross each other at the Fermi level when $E_{\alpha +} = E_{\beta -} = 0$, which requires 
  \begin{eqnarray}
    \cos k^0_{x \mp} &=& \frac{\sqrt{(\mp 2t_2-\delta+N_{\alpha}-\mu)^{2} - (\Delta_{\alpha})^{2}}}{2(t_1\pm 2t_3)} \nonumber\\
    &=& \mp \frac{\sqrt{(\mp 2t_1+\delta+N_{\beta}-\mu)^{2} - (\Delta_{\beta})^{2}}}{2(t_2 \pm 2t_3)}.
    \end{eqnarray}
Here, $ k^0_{x -}$ and $ k^0_{x +}$ give the locations of DPs associated with the Dirac cones $D_1$ and $D_2$ along $\Gamma$-X and ${\rm X^{\prime}}$-M directions, respectively. It may also noted that $ k^0_{y -} = 0$ and $ k^0_{y +} = \pi$. 

\subsection{Linearized dispersion}
Using Taylor expansions of $E^x_{\alpha +}$ and $E^x_{\beta -}$ around DPs along $k_x$ with the help of Eq. (9), one obtains linearized dispersion $E^x_{s \mp} = c^x_{s \mp}q_x$, where the constant $c^{x}_{s \mp}$ is given by 
  \begin{eqnarray}
    c^{x}_{\alpha \mp} = \mp \frac{2(t_1 \pm 2t_3)^2}{|\mp 2t_2 - \delta + N_{\alpha} - \mu|} \sin 2k_x^{0} \nonumber \\
    c^{x}_{\beta \mp} = \pm \frac{2(t_2 \pm 2t_3)^2}{|\mp 2t_1 + \delta + N_{\beta} - \mu|} \sin 2k_x^{0}.
    \end{eqnarray}
Note that the subscript $ -$ and $+$ in $c^{x}_{\alpha \mp}$ refers to the DPs $D_1$ and $D_2$, respectively. 

$H_{\rm MF}$ is not in the block-diagonal form along $k_y$-direction. Moreover, the Hamiltonian for $\uparrow$-spin electron is two-fold degenerate at the DPs. Therefore, the degenerate perturbation theory yields the following corrections to the energies near the DPs along $k_y$
    \begin{equation}
    E^y_{{\pm}} =  \pm  \left(\frac{\Delta_{\alpha} - e_{\alpha}}{b_{\alpha}} + \frac{\Delta_{\beta} + e_{\beta}}{b_{\beta}}\right) \epsilon^{\alpha \beta}_{xy} 
    \end{equation} 
when $\epsilon^{\alpha \beta}_{xy}$ is very small. Note that the subscript $s$ has been dropped here because the band along $k_y$ is not far away from an even mixture of both the orbitals. $ b_{\alpha} = \epsilon^{\alpha \alpha}_x + \epsilon^{\alpha \alpha}_{xy} $ and $e_{\alpha} = \sqrt{b_{\alpha}^2 + \Delta_{\alpha}^{2}}$. The linear dependence of the energy dispersion near the DP is readily obtained from $\epsilon^{\alpha \beta}_{xy}$, where $\sin q_y$ can be approximated by $q_y$ for small $q_y$ so that $E^{y}_{s\pm} = d_{s\pm} q_y$. $d_{s\pm}$ is a constant given by 
\begin{equation}
 d_{s\pm} = \pm 4t_4 \sin k_x^{0}(f_{\alpha \mp} + f_{\beta \mp}),
\end{equation}
where
\begin{eqnarray}
f_{\alpha \mp} &=& \frac{\Delta_{\alpha} - |\mp 2t_2-\delta+N_{\alpha}-\mu|}{\sqrt{(\mp 2t_2-\delta+N_{\alpha}- \mu)^2-\Delta^2_{\alpha}}} \nonumber \\
f_{\beta \mp} &=& \frac{\Delta_{\beta} + |\mp 2t_1+\delta+N_{\beta}-\mu|}{\sqrt{(\mp 2t_1+\delta+N_{\beta}- \mu)^2-\Delta^2_{\beta}}} .
\end{eqnarray}
Upper and lower sign correspond to the Dirac cones $D_1$ and $D_2$, respectively. Eqs. 11-15 provide the conditions for the coexistence of Dirac semimetallic and SDW state as well as the linear energy dispersion in the vicinity of DPs. 

\begin{figure}
 \includegraphics[scale=1.0, width=9.2cm]{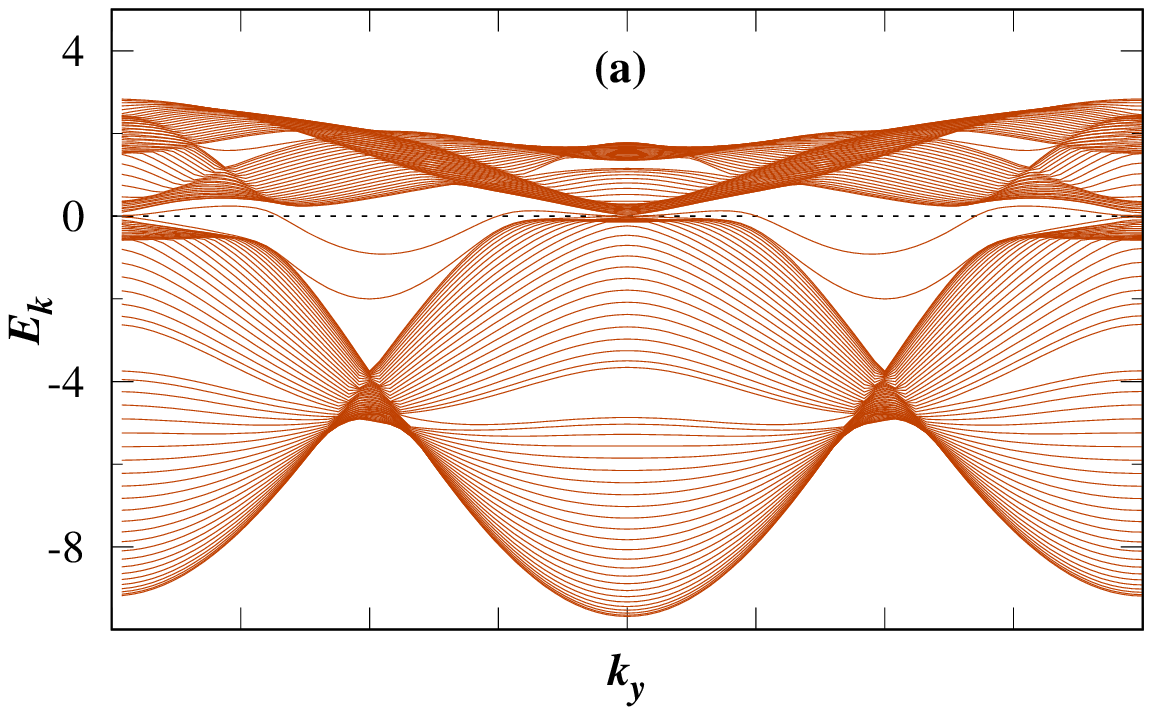}
 
 \vspace{-6mm}
 
 \includegraphics[scale=1.0, width=9.2cm]{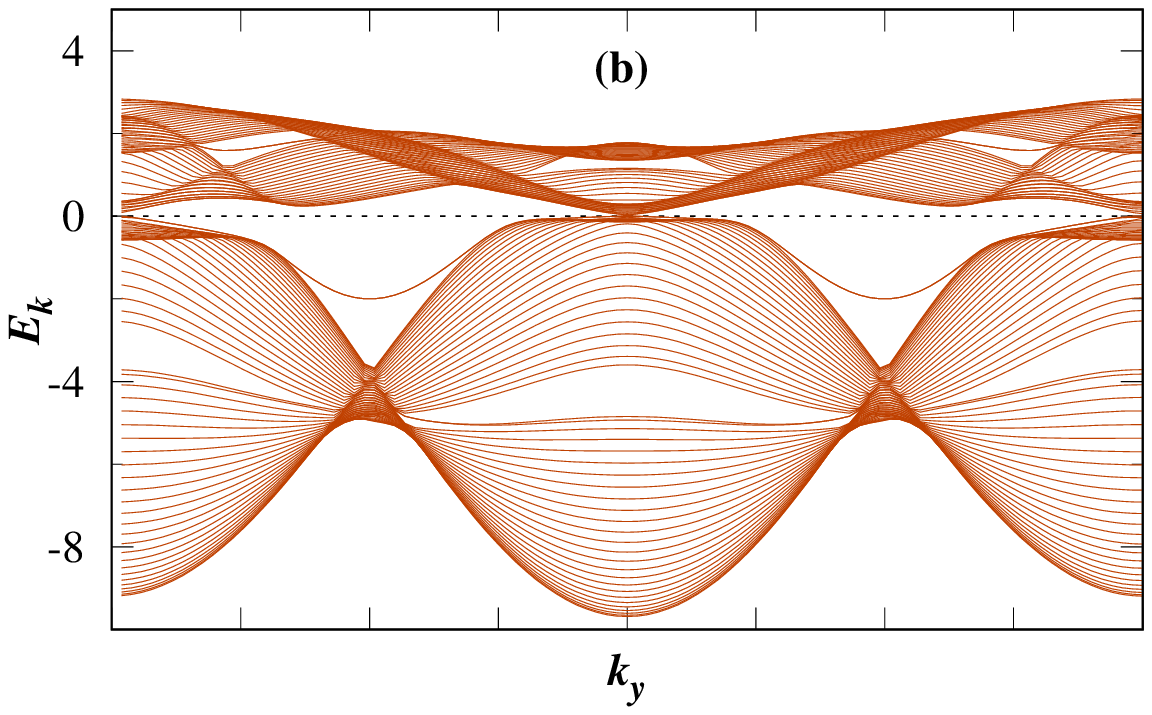} 
  
  \vspace{-6mm} 
  \includegraphics[scale=1.0, width=9.2cm]{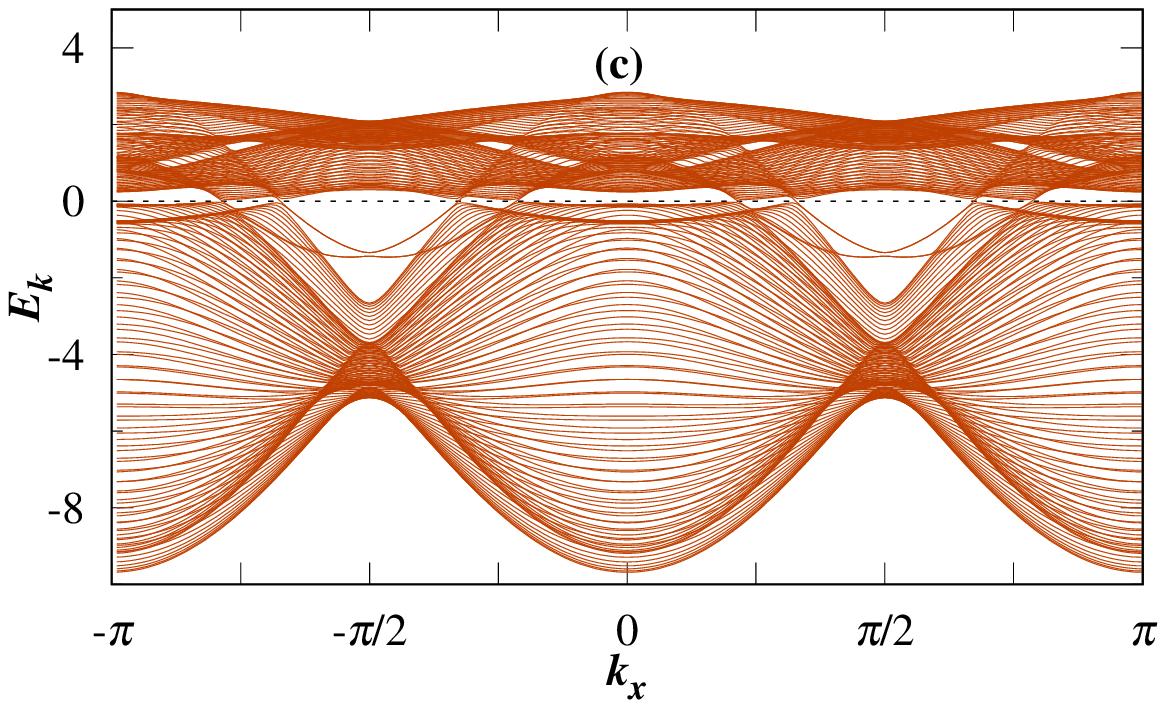}
   \caption{Energy bands for the DS-SDW state for a ribbon of width (a) $W$ = 50, (b) $W$ = 51 considered along $x$-direction and (c) $W$ = 50 considered along $y$-direction for $U = 4$ and $J = 0.18U$, for which, the orbital charge densities are  $n_{\alpha} = 1.14$ and $n_{\beta} = 0.86$ whereas the magnetic-exchange fields are $\Delta_{\alpha} = 0.58$ and $\Delta_{\beta} = 0.53$.}
    \label{edge}
\end{figure} 

\subsection{Edge states} An edge state may exhibit behavior which is different from the bulk band as in the case of topological insulator where the bulk bands are gapped while charge transport occurs by the topologically protected surface states~\cite{cai}. In iron pnictides, there exists localized edge states in the high temperature phase even without long-range order~\cite{lau}. It may be noted that these edge states, nearly degenerate, are not associated with any Dirac cones, which are absent in the paramagnetic state. For a given $k_y$, these edge states are bonding and antibonding mixture of states localized at the edges of ribbon/strip extending along $y$ direction. They result from the fact that a one-dimensional Hamiltonian for a given $k_y$ can be deformed continuously to one which has topologically protected winding number, while the edges states are preserved in the deformation process.

Next, we examine the edge states in the DSM-SDW state with ribbons oriented either along $x$ or $y$ directions. First, a ribbon of width $W$ ($W$ atomic sites) lying along $y$ direction is considered so that $k_y$ is a good quantum number. The ribbon Hamiltonian with dimension $2W \times 2W$ is given by
 \begin{equation}
H_{Rbx}({\bf k}) = 
\begin{pmatrix}
H_+  &  H^{\prime} & O &\cdots \\
H^{\prime \dagger}  &  H_- & H^{\prime} &\cdots \\
O & H^{\prime \dagger} & H_+ & \cdots\\
\vdots & \vdots & \vdots & \ddots
\end{pmatrix}
\end{equation}
where
\begin{equation*}
H_{\pm} = 
\begin{pmatrix}
\epsilon^{\alpha \alpha}_y \mp \Delta_{\alpha} - \delta + N_{\alpha}  - \mu  &  0 \\
0  &  \epsilon^{\beta \beta}_{y} \mp \Delta_{\beta} + \delta + N_{\beta} - \mu
\end{pmatrix}
\end{equation*}

\noindent and

\begin{equation*}
    H^{\prime} = 
    \begin{pmatrix}
    -t_1 - 2t_3 \cos k_y  &   -2it_4 \sin k_y \\
    -2it_4 \sin k_y  &    -t_2 - 2t_3 \cos k_y
    \end{pmatrix}.
\end{equation*} 
 Similarly, when the ribbon's length is oriented along $x$ direction so that $k_x$ is a good quantum number, the Hamiltonian $H_{Rby}({\bf k})$ with size $4W \times 4W$ has a form similar to that of $H_{Rbx}$.
 However, $H_{\pm} $ and $H^{\prime}$ have size $4 \times 4$ instead. These matrices are given by
 \begin{equation*}
     H_+ = H_- = H= 
    \begin{pmatrix}
    h_{AA}  &  h_{AB}  \\
    h_{BA}    &  h_{BB} 
    \end{pmatrix}
\end{equation*}
with
\begin{equation*}
    h_{AA}({\bf k}) = 
    \begin{pmatrix}
    -\Delta_{\alpha} - \delta + N_{\alpha} - \mu  &  0 \\
      0   &  -\Delta_{\beta} + \delta + N_{\beta} - \mu
    \end{pmatrix}
\end{equation*}   
and
\begin{equation*}
    h_{AB}({\bf k}) = 
    \begin{pmatrix}
    \epsilon^{\alpha \alpha}_x &  0 \\
      0   &  \epsilon^{\beta \beta}_x
    \end{pmatrix}
\end{equation*} 
\noindent while
\begin{equation*}
    H^{\prime}({\bf k}) = 
    \begin{pmatrix}
      -t_2  &  0  &  - 2t_3 \cos k_x  &   -2it_4 \sin k_x \\  
      0  &  -t_1  &   -2it_4 \sin k_x  &   -2t_3 \cos k_x \\ 
      - 2t_3 \cos k_x   &  -2it_4 \sin k_x  &  - t_2  &   0 \\
        -2it_4 \sin k_x  &   -2t_3 \cos k_x &  0  &   -t_1 
    \end{pmatrix}.
\end{equation*}

There exists nearly degenerate two edge states for the ribbon in the paramagnetic state which may or may get clearly separated in the presence of magnetic order depending on whether $W$ is odd or even. Fig. \ref{edge} show the edge state dispersions in the DS-SDW state. There are two edge states localized at the edges of ribbon when $W$ is even. Another partially visible edge-state like  dispersion stands out clearly from the bulk bands. These edge states in the DS-SDW state are dispersing unlike those in graphene, which are flat~\cite{bernevig}. One of the edge-state dispersion crosses the Fermi level while the other one not. The clear separation of edge dispersions is missing when $W$ is an odd, which is simply a consequence of the same magnetic-exchange fields at both the edges. It may be noted that they are always degenerate only at $k_x = 0$ and $k_y = \pm \pi$. Fig. \ref{edge}(c) shows the dispersion in a ribbon oriented along the $x$ direction. As expected, they are symmetric about $k_x = -\pi/2$ an $\pi/2$ because of the two-sublattice structure along the ribbon extending infinitely along $x$. None of the edge-state dispersion is flat and both can be noticed to cross the Fermi surface.

\begin{figure}
    \centering
    \includegraphics[scale=1.0, width=9.3cm]{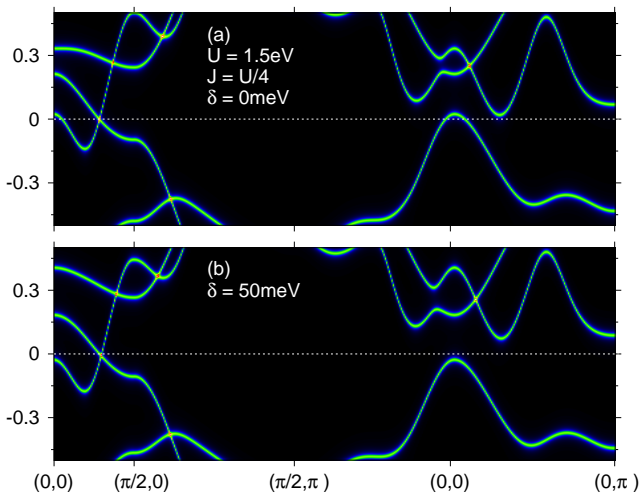}
    \vspace{-0mm}
    \caption{Electron dispersion in the $(\pi, 0)$ SDW state of five-orbital model of Graser \textit{et. al.} when (a) $\delta = 0$ (b) $\delta = 50$meV, where $U = 1.5eV$ and $J = 0.25U$. The hole pocket at $\Gamma$ disappears in the presence of OS $\delta$ =50meV.}
    \label{5}
\end{figure} 

\section{DS-SDW state in five-orbital model}
Finally, we discuss DS-SDW state in a five-orbital model which describes the band structure more realistically. We consider the five-orbital model due to Graser \textit{et. al.}. In the unordered state, it has similarities as well as differences from the two-orbital model. There is a hole pocket around $\Gamma$ and an electron pocket around X while the hole pocket around M is absent. The electron pocket is mostly dominated by $d_{xy}$ orbital~\cite{raghu,graser}. 

Fig.~\ref{5} shows the electronic dispersion in a self-consistently obtained SDW state for $U = 1.5$eV and $J = 0.25U$ with and without orbital splitting. The ratio $U/W \sim$ 0.25 with $W\sim6$eV being bandwidth is in accordance with various estimates~\cite{yang}. The OS parameter $\delta$ is taken to be 50meV, which is nearly of similar order in observed in the experiments. The evidence of the splitting already present in the high-temperature phase comes from transport~\cite{blomberg}, ARPES~\cite{yi1,yi2} and magneto-torque measurements~\cite{kasahara1}. 

It may be noted that the unlike the two-orbital model, there is a large hole pocket around $\Gamma$ and a tiny hole pocket located not far away from $\Gamma$ along $\Gamma$-X. With increasing OS parameter $\delta$, which lowers the energy of $d_{xz}$ orbital, the $d_{xz}$-dominated bands such as the hole pocket around $\Gamma$ will be pushed down below the Fermi level as shown in Fig. \ref{5} (a) and (b). There are two important differences from two-orbital models with regard to the Dirac cone. First, the difference in the magnitude of slopes of the crossing bands at DPs are not as large as in the two-orbital model. Second, along $k_x$, one of the crossing bands is largely dominated by $d_{xz}$ while the other one by $d_{xy}$ orbital unlike the $d_{yz}$ orbital. Therefore, when $\delta$ is changed, the corresponding shift in the positions of DPs is relatively small in comparison to the two-orbital model. However, the DPs are already very close to the FS even in the absence of OS. Thus, the important role of OS $\delta$ is to suppress the hole pocket around $\Gamma$ point, which is necessary to obtain the DS-SDW state.

\section{Summary and conclusions} 
In the two-orbital model, the pair of Dirac cones in the SDW state are located away from the Fermi surface. While one pair is above, the other one is below. The separation of these pairs with respect to the Fermi surface can be minimized with the help of OS. Inclusion of OS pushes up the Dirac cone found below the Fermi level and pushes down the one above it, which is possible because of a sharp difference in the slopes of the crossing bands dominated largely by a single orbital. For a given OS parameter, an overall small shift away from the Fermi surface may also occur despite both pairs of DPs being at the same energy level. However, such a shift can be overcome by either doping holes or electrons.  The signature of shifting of DPs can be observed through various experiments such as transport measurements, quantum oscillation, ARPES etc.~\cite{blomberg}.

In the current work, a detailed study of the DS-SDW state was carried out in the two-orbital model because of its simplicity, as it allows for obtaining conditions for the DS-SDW state in an analytical form. However, DS-SDW state can also exist in a more realistic five-orbital model in a region of interaction and OS parameters space, which is illustrated for such a particular set within the range in accordance with various estimates. However, there are several important differences from the two-orbital model in terms of orbital content of the Dirac cone and presence of hole pocket around $\Gamma$ points. Another major difference exists in terms of the number of pairs of Dirac cones. There is only one pair of Dirac cones close to the Fermi level in the five-orbital model~\cite{richard}, for which, it may relatively be easier to bring the DPs at the Fermi level. Overall, we find that, the OS pushes the hole pocket around $\Gamma$ below and secondly it may also shift the DPs. As a result, it is not only possible to control electronic properties including the charge transport by tuning the OS, thus, in turn, regulate the contribution of Dirac cone but it is also feasible to realize DS-SDW state.  

The OS parameter plays a crucial role in obtaining the DS-SDW state, whose inclusion here has been motivated by its presence in the high-temperature state with tetragonal symmetry and nematic order~\cite{watson,vojta,hinkov,baek}. 
The nematic phase is marked by the presence of a short-range order with broken four-fold rotation symmetry~\cite{kasahara} and is observed at a temperature $T > T_{SDW}$ in several IBSs~\cite{yamase}.
It was noted that the lattice anisotropy in the orthorhombic symmetry cannot give rise to a splitting as large as $\approx$ 60meV between the $d_{xz}$ and $d_{yz}$ orbitals. This perhaps indicates that the OS, which persists into the high temperature tetragonal phase, in addition to the orthorhombic symmetry or magnetic order, could be the key factor behind the anisotropic electronic behavior. Therefore, there is no reason why such a splitting should be ignored in the low-temperature phases such as the SDW state. Although an additional splitting will also be generated by the SDW state itself as it breaks the four-fold rotation symmetry. Experimentally, OS can be controlled by applying appropriate pressure such as an in-plane stress~\cite{chu}. 

In conclusions, we have examined the possible existence of spin-density wave state with Dirac points at the Fermi level in iron pnictides such that the state is essentially a Dirac semimetal without any other band crossings. We find that such a state indeed can be obtained in the presence of finite OS. While the interaction parameters are not tunable experimentally, charge carrier doping may shift entire bands up or down, the OS of otherwise degenerate $d_{xz}$ and $d_{yz}$ orbitals, on the other hand, can suppress other bands crossing the Fermi level while leaving the Dirac points in the vicinity of Fermi surface. Thus, the OS, tunable experimentally with the help of in-plane mechanical stress, can be used to modify the electronic states near the Fermi surface in order to control the electronic properties as well as in obtaining DS-SDW state. A semimetallic state with magnetic order considered here can be used to explore low-energy collective excitations to gain insight into the role of Dirac cones and nodes in correlated systems. 

\section*{ACKNOWLEDGEMENTS} 
D.K.S. was supported through DST/NSM/R\&D\_HPC\_Applications/2021/14 funded by DST-NSM and start-up research grant SRG/2020/002144 funded by DST-SERB. D.K.S. greatfully acknowledges M. Biderang, A. Akbari, S. K. Mahatha and Y. Bang for useful discussions.

\end{document}